\documentclass[a4paper]{jpconf}
\usepackage{graphicx}
\usepackage{paralist}
\begin{document}
\title{ALICE HLT TPC Tracking of Pb-Pb Events on GPUs}

\author{David Rohr$^1$, Sergey Gorbunov$^1$, Artur Szostak$^2$, Matthias Kretz$^1$, Thorsten Kollegger$^1$, Timo Breitner$^1$, Torsten Alt$^1$ for the ALICE HLT Collaboration}

\address{$^1$~Johann-Wolfgang-Goethe University, Frankfurt, Germany\\ $^2$~Department of Physics and Technology, University of Bergen, Norway}

\ead{drohr@cern.ch}

\begin{abstract}
The online event reconstruction for the ALICE experiment at CERN requires processing capabilities to process central Pb-Pb collisions at a rate of more than~200 Hz, corresponding to an input data rate of about~25~GB/s.
The reconstruction of particle trajectories in the Time Projection Chamber (TPC) is the most compute intensive step.
The TPC online tracker implementation combines the principle of the cellular automaton and the Kalman filter.
It has been accelerated by the usage of graphics cards (GPUs). A pipelined processing allows to perform the tracking on the GPU, the data transfer, and the preprocessing on the CPU in parallel.
In order for CPU pre- and postprocessing to keep step with the GPU the pipeline uses multiple threads.
A splitting of the tracking in multiple phases searching for short local track segments first improves data locality and makes the algorithm suited to run on a GPU.
Due to special optimizations this course of action is not second to a global approach.
Because of non-associative floating-point arithmetic a binary comparison of GPU and CPU tracker is infeasible.
A track by track and cluster by cluster comparison shows a concordance of 99.999\%.
With current hardware, the GPU tracker outperforms the CPU version by about a factor of three leaving the processor still available for other tasks.
\end{abstract}

\section{Introduction}

ALICE (A Large Ion Collider Experiment) is a dedicated Pb-Pb detector designed to exploit the physics potential of nucleus-nucleus interactions at the Large Hadron Collider at CERN~\cite{alice_technical-proposal,alice_technical-paper}.

The base-line design consists (from inside out) of a high-resolution Inner Tracking System (ITS), a cylindrical Time Projection Chamber detector (TPC), particle identification detectors, and a single-arm electromagnetic calorimeter.

\begin{figure}[ht]
\begin{minipage}{18pc}
\includegraphics[width=18pc]{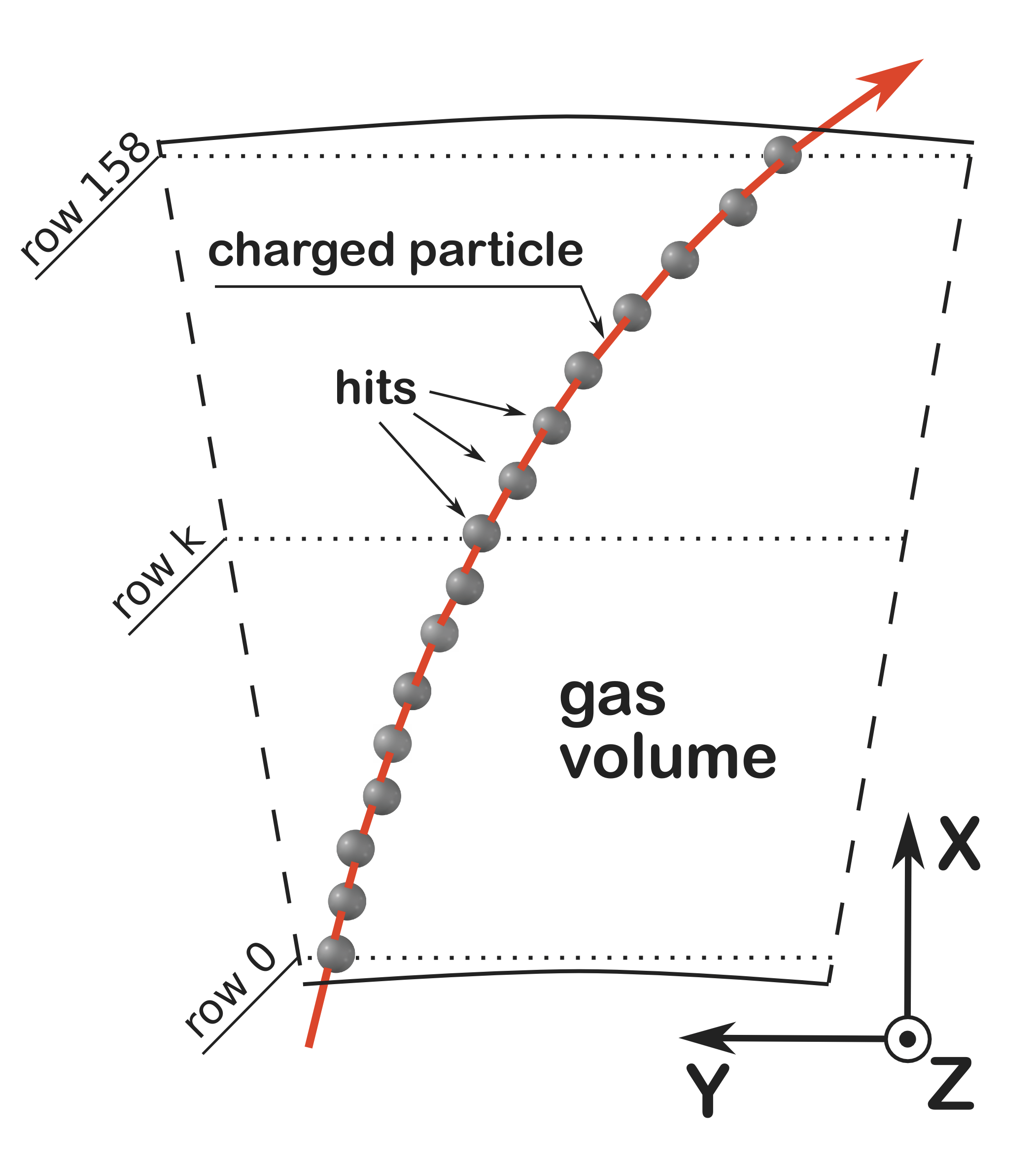}
\caption{\label{fig:HLTSectorGeometry}Geometry of a TPC sector.}
\end{minipage}\hspace{2pc}%
\begin{minipage}{18pc}
\includegraphics[width=18pc]{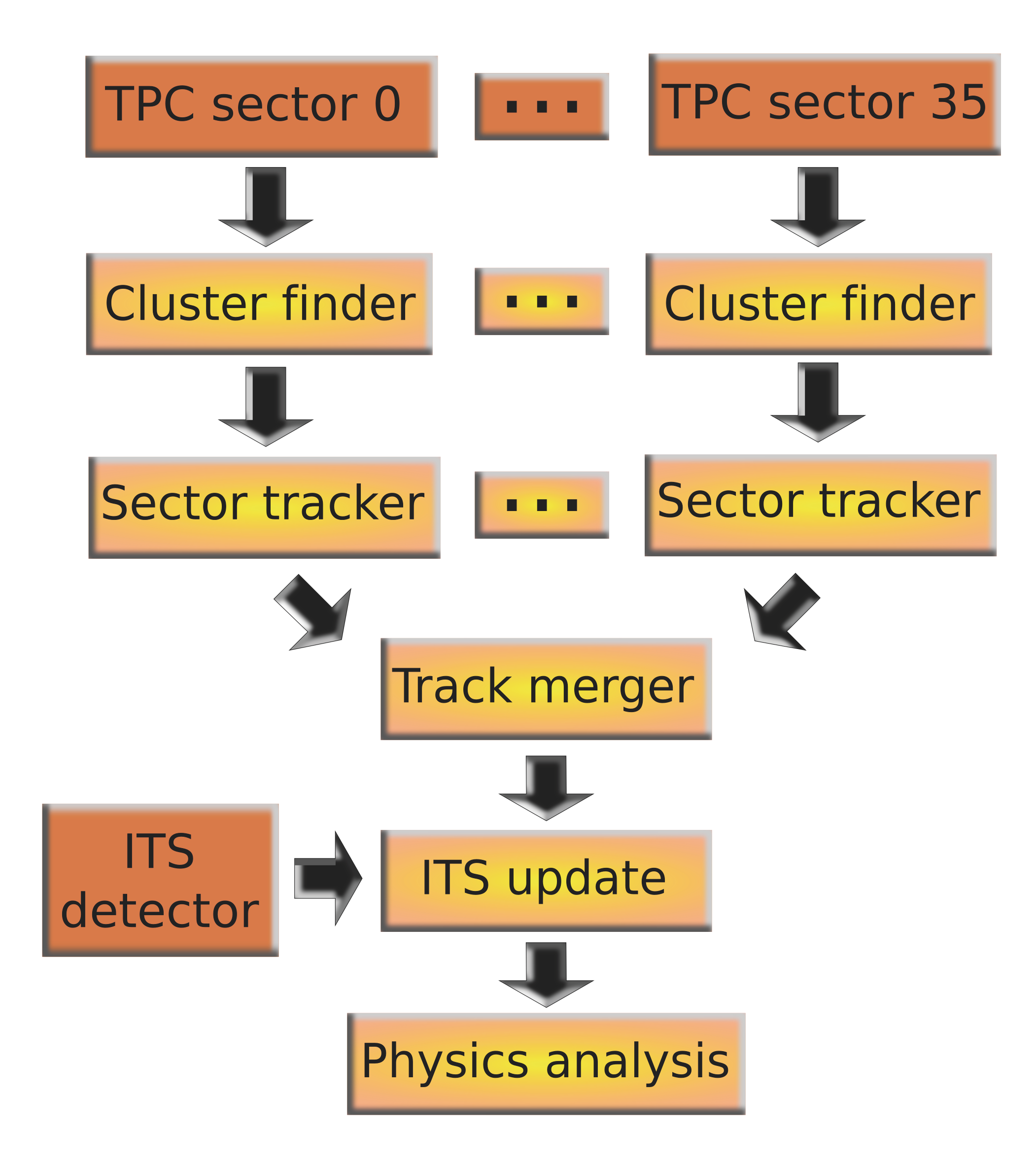}
\caption{\label{fig:HLTReconstructionScheme}HLT reconstruction scheme.}
\end{minipage}
\end{figure}

Reconstruction of the particle trajectories (tracking) is one important step during event reconstruction.
The Time Projection Chamber (TPC) detector is the main tracking detector of ALICE.
Traversing particles ionize gas molecules inside the TPC.
The ionization points are measured and are called clusters.
Computing the TPC tracks from the clusters is a major part of the event reconstruction and it is computationally very expensive.
The TPC detector consists of two cylindrical volumes placed along the beam; either volume is split into~$18$ trapezoidal readout sectors.
The detector measures track positions on 159 rows as it is shown in Fig.~\ref{fig:HLTSectorGeometry}.

The online event reconstruction in ALICE is performed by the High-Level Trigger. The overall online reconstruction scheme is presented in Fig.~\ref{fig:HLTReconstructionScheme}. It starts with the TPC cluster finder, which identifies clusters out of the TPC raw data. These reconstructed clusters are sent to the sector tracker which reconstructs the tracks in each TPC sector individually. Then the sector tracks are merged by the track merger algorithm, and later updated with the measurements from the ITS detector. The reconstruction of the event's vertex and the physical triggers are run at the end of the reconstruction tree structure. Typically, every processing stage
reduces the size of the event data. This scheme processes data as early as possible avoiding any unnecessary copy steps and uses all available data locality and parallelization.

The core of the event reconstruction takes place in the TPC sector tracker, which creates tracks from the TPC measurements.
It is the only component which processes the TPC clusters, the higher level components operate on the reconstructed sector tracks.

\begin{figure}[ht]
\includegraphics[width=21pc]{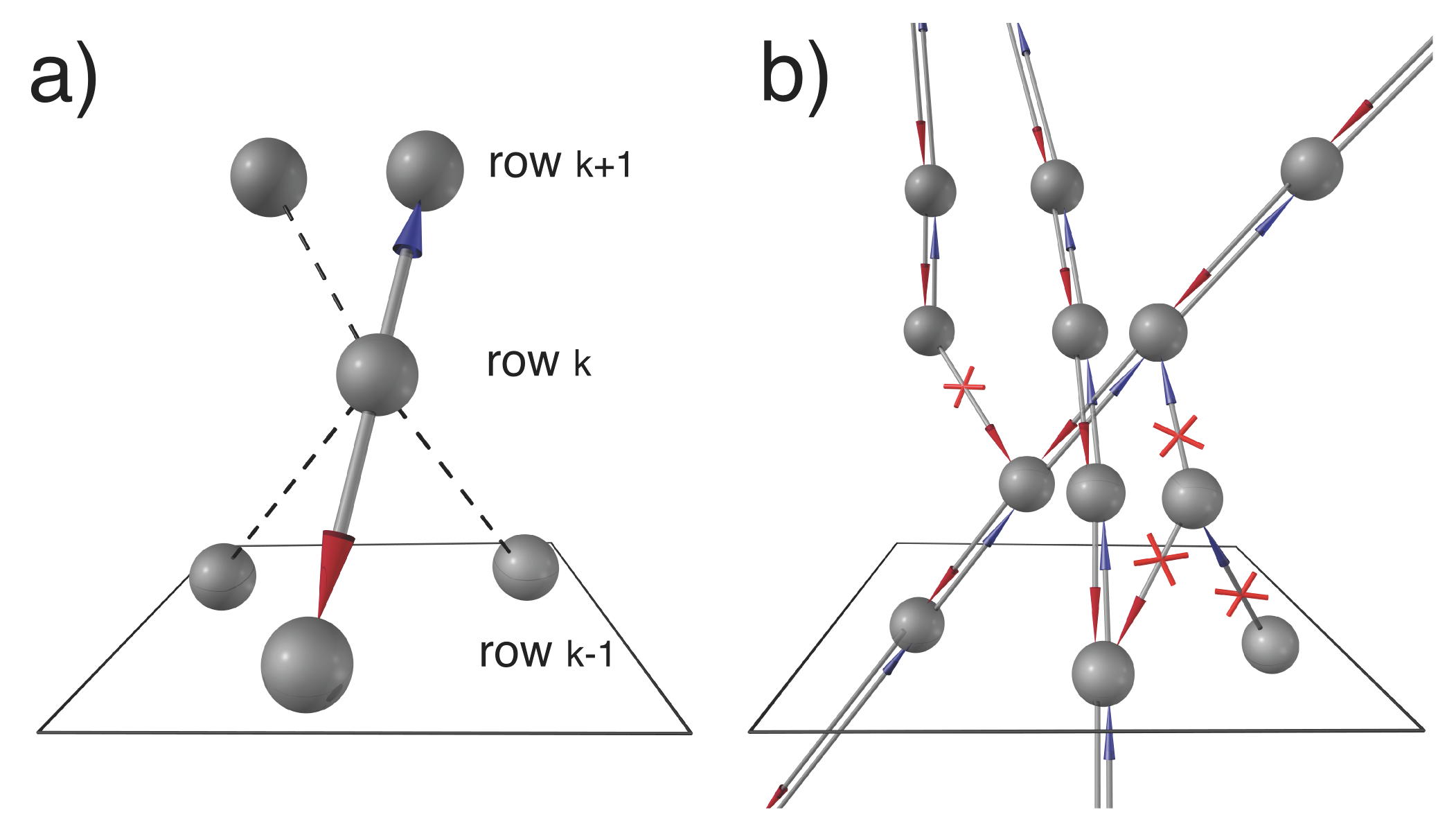}\hspace{2pc}\newline
\begin{minipage}[b]{30pc}\caption{\label{fig:HLTSteps}a) Neighbors finder. b) Evolution step of the Cellular Automaton.}
\end{minipage}
\end{figure}

The tracking  algorithm starts with a combinatorial search for track candidates
(tracklets), which is based on the Cellular Automaton method~\cite{CARef}.
Local parts of trajectories are created from geometrically nearby clusters, thus eliminating unphysical cluster combinations at the local level. The combinatorial processing consists of the
following two steps:
\begin{itemize}
\item Neighbors finder: For each cluster at a row k the best pair of neighboring clusters from
rows k+1 and k-1 is searched for, as it is shown in Fig.~\ref{fig:HLTSteps}a. The neighbor selection criterion
requires the cluster and its two best neighbors to form a straight line. The links to the best
two neighbors are stored. Once the best pair of neighbors is found for each cluster, the
step is completed.

\item Evolution step: Reciprocal links are determined and saved, all the other links are
removed (see Fig.~\ref{fig:HLTSteps}b).
\end{itemize}

Every saved one-to-one link defines a part of the trajectory between the two
neighboring clusters. Chains of consecutive one-to-one links define the tracklets.
After the tracklets are created, the following steps are executed:

\begin{itemize}
\item Tracklet construction: Track parameters are fit to the tracklets using the Kalman filter. These parameters are then used to extrapolate the trajectory to adjacent rows and search for more
clusters belonging to the tracklet. When meeting a $\chi^2$-condition the clusters are collected to the tracklet and the track parameters are refitted. This is repeated until no more clusters are found.

\item Tracklet selection: Some of the track candidates can have overlapping parts. In this
case the longest track is saved, the shortest removed. A final quality check is applied to
the reconstructed tracks, including a cut on the minimal number of clusters and a cut for low
momentum.
\end{itemize}

\section{Sector Tracking on GPU}

The original ALICE HLT TPC tracker was designed with parallelism in mind. Within all tracking steps multiple tasks can be executed simultaneously, e.g.~producing the links in the neighbors finder can be done for each cluster independently or extrapolation and fitting of different tracklets in the tracklet constructor can be done in parallel. The GPU tracker uses one thread per cluster or per track in each of the steps. The implementation is such that GPU and CPU tracker share most of the source code in a common file. Only small special wrappers exist for both particular architectures that include the common files. This approach greatly improves the maintainability.

Besides the steps of the above-described algorithm its implementation adds two more tasks: initialization and track output. These steps merely perform data reformatting and are memory bound while touching most bytes only once. Thus, due to limited PCIe bandwidth these tasks cannot benefit from the GPU and are still processed by the CPU. For all other steps a GPU implementation was developed.

\begin{figure}[ht]
\includegraphics[width=35pc]{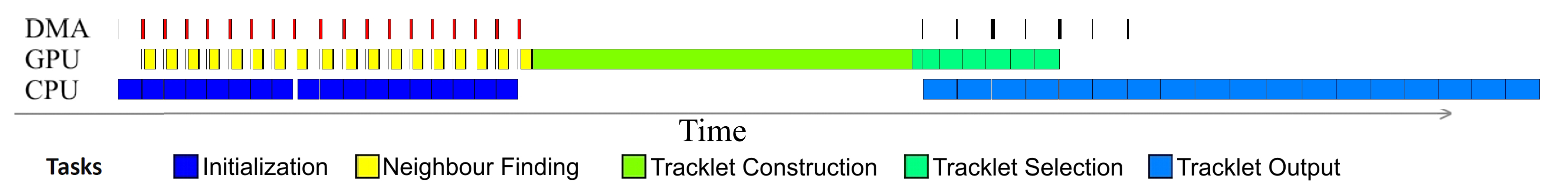}\hspace{2pc}\newline
\begin{minipage}[b]{30pc}\caption{\label{fig:async_oldout}Pipeline of the first Fermi Tracker Implementation.}
\end{minipage}
\end{figure}

For ensuring good GPU utilization, processing of the sectors is arranged in a pipeline such that while the GPU performs the tracking for sector~$i$ the CPU can preprocess sector~$i - 1$. During development and operation of the GPU tracker, new processors and GPUs became available. Initially, the GPU tracker ran on Intel Nehalem quad-core CPUs and NVIDIA GTX~295 GPUs. Now, the GPU tracker runs on AMD Magny-Cours twelve-core processors and GTX480 GPUs. The new processor has a higher peak performance considering all twelve cores, but for single thread applications the old processor is faster. Since the original pipeline of the GPU tracker only used a single thread, the hardware update sped up the tracking but slowed down pre- and postprocessing. In addition, a change to the output format slowed down the postprocessing even more. As a consequence, the single-threaded pipeline that works well on the old hardware is unsuited for the new compute nodes. Fig.~\ref{fig:async_oldout} shows the measured times of the pipeline steps of the first Fermi tracker version.

\begin{figure}[ht]
\includegraphics[width=35pc]{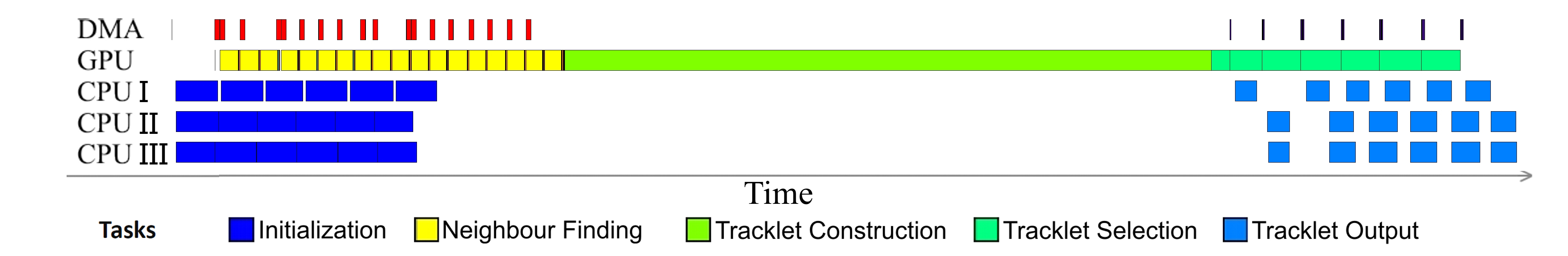}\hspace{2pc}\newline
\begin{minipage}[b]{30pc}\caption{\label{fig:async_multithread}Pipeline of the Fermi Tracker with Multithreading.}
\end{minipage}
\end{figure}

The figure reveals that the GPU is idling for a significant amount of time. Two optimizations approach this problem. At first, algorithmic optimizations to the output procedure could reduce the computation time. Second, a multithreaded version of the pipeline was developed, that uses multiple CPU cores which alternately process the tasks in the pipeline. The pipeline processing is visualized in Fig.~\ref{fig:async_multithread}. The GPU remains busy, except for the initialization of the first and the output of the last sector in the pipeline.

\section{Consistency of GPU and CPU Tracking}

The functionality of both CPU and GPU tracker is examined with Monte-Carlo simulations. Both trackers run on the set of clusters produced in the simulation. Afterward, the obtained tracks are compared with the reference tracks from the simulation. Under this perspective tracking efficiencies of GPU and CPU tracker show no difference. Another possibility to compare the results is to compare physical statistics derived from the tracks. An analysis of various statistical quantities revealed that there is only one which shows a slight variation: the number of clusters per track. The reason is that both trackers find essentially the same tracks but in some cases assign the clusters differently.

Although GPU and CPU tracker results are basically equivalent, it is desirable to do a comparison as direct as possible - in the best case on a bit level. However, such a bitwise comparison is impossible for three reasons:
\begin{compactitem}
\item Indeterministic cluster to track assignment,
\item Inconsistent sorting of the tracks,
\item Non-associative floating-point arithmetic.
\end{compactitem}

\subsection{Cluster assignment}

The initial rule for the assignment was to assign the clusters to the longest track possible. In the case of two tracks of the same length the first one was used. Because of the multithreading of the GPU tracker the order in which the tracklets are constructed is not well defined. Therefore, the cluster assignment for two tracks of identical length is not deterministic. The GPU and CPU tracks themselves are almost identical but in rare cases a cluster is assigned to another track. This made the results incomparable.

A better criterion for the cluster assignment has thus been searched for. The idea is to use a continuous measure instead of the discrete tracklet length. It is self-evident that the $\chi^2$ value (normalized residual between the clusters and the trajectory) can be used for this purpose. Multiple possibilities exist:
\begin{compactitem}
\item{} Using the residual between the cluster and the track.
\item{} Using the residual of the entire track.
\item{} Using a combination of the residual and the track length.
\end{compactitem}
The first option naively sounds like a good idea. However, the results are merely bad. Usually at least two tracklets for one track are found. This track is then reconstructed twice resulting in two tracklets with almost identical clusters. It is desired to keep the better tracklet, assign all clusters to this tracklet, and remove the other instance of the track as a clone. However, when using the residual between the cluster and the track, about halve of the clusters get assigned to the one instance while the rest is considered belonging to the other instance. The clone is not removed. Therefore, the residual for the full track is used.

\begin{figure}[ht]
\includegraphics[width=30pc]{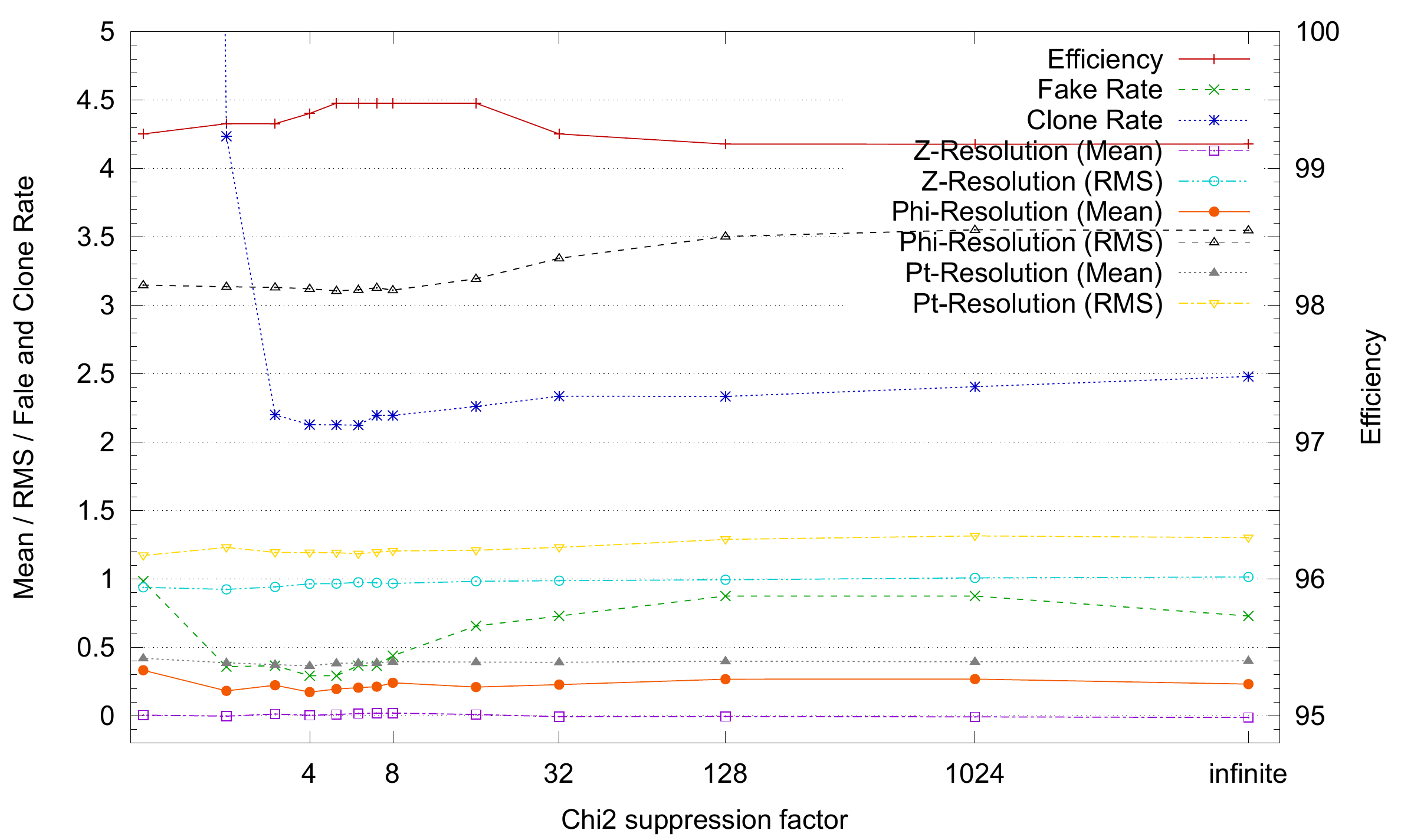}\hspace{2pc}\newline
\begin{minipage}[b]{30pc}\caption{\label{fig:gp_chi2}Efficiency and Resolution using different $\chi^2$ Suppression Factors.}
\end{minipage}
\end{figure}

Using only the residual for the assignment leads to unsolvable problems with short tracks, as it is often simple to fit a trajectory to only few clusters. Thus, a combination of the tracklet length and the residual is better. For accomplishing this, a cluster weight~$w$ is implemented as $w = n \cdot (\alpha - \frac{\chi^2}{\beta})$. The value of~$\beta$ is chosen such that $\frac{\chi^2}{\beta}$ is of order~$1$ for normalization reasons. The factor~$\alpha$ is called the~$\chi^2$-suppression factor and~$n$ is the length of the tracklet. The clusters are then assigned to the tracks such that the cluster weights are maximum. Thus,~$\alpha = \infty$ results in the old behavior where only the tracklet length is decisive. In general, the bigger~$\alpha$ the lower is the influence of~$\chi^2$. Fig.~\ref{fig:gp_chi2} shows how the tracking quality varies with~$\alpha$.

It turns out that incorporating the residual even improves the tracking efficiencies, reduces clone and fake rates, and either improves or maintains the resolutions. For very small~$\alpha$ the tracking gets unstable. Finally, in the tracker a value of~$\alpha = 6$ is employed as it is considered the best tradeoff between efficiency and stability.

\subsection{Track Sorting}

Analogously to the tracklet order which led to an indeterministic cluster assignment as described above, the order of the final sector tracks after the tracklet selection can have a marginal effect on the track merging. This is overcome by a fast sorting of the tracks in between of sector tracking and track merging.

\subsection{Floating-Point Arithmetic}

With both above problems solved, no other effect of concurrency on the tracking result is observed. Still, different compilers (or even different compiler options) produce results which are not bitwisely identical because the floating-point arithmetic is not associative leading to different rounding. Unfortunately, there is no way to negate this but the effect is mitigated by the design of the tracker. The sector tracker only performs the track finding. It outputs a list of clusters for each track and the track merger does a refit of the track. Hence, as long as despite of different rounding the same clusters are found, the result will be bitwisely the same because the varying intermediate calculations are not used by the merger. If within one row two clusters are equally close to the trajectory, depending on the rounding either the one or the other can get assigned to the track. In such cases the result differs.

\begin{figure}[ht]
\includegraphics[width=26pc]{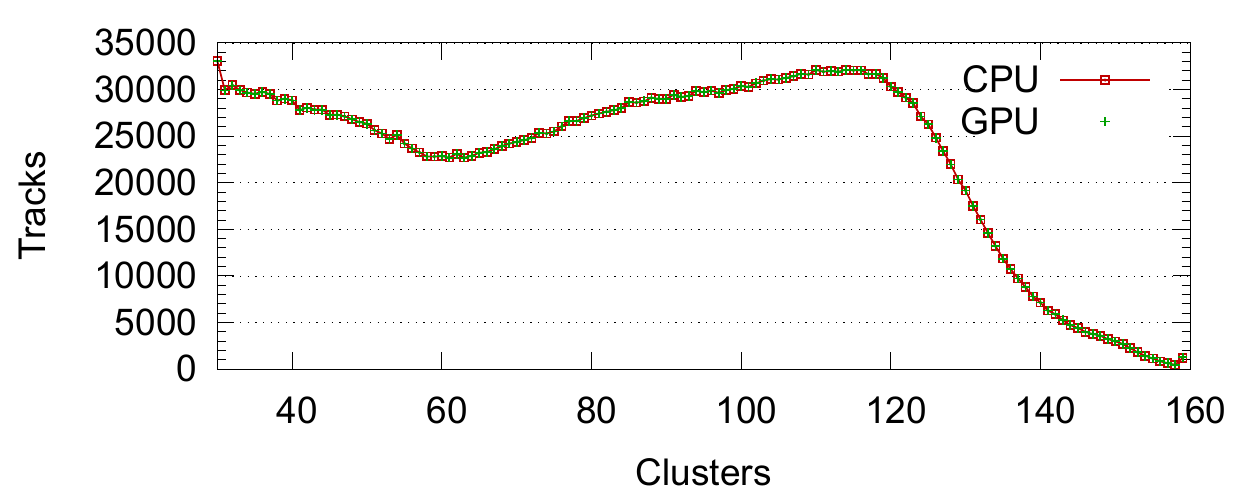}\hspace{2pc}\newline
\begin{minipage}[b]{30pc}\caption{\label{fig:gp_sizesum}Comparison of GPU and CPU Cluster per Track Statistics after the Improvements.}
\end{minipage}
\end{figure}

The influence on the tracking result is analyzed in the following way: A sample of events is processed by CPU and by GPU. A one-to-one correspondence of the resulting tracks is determined. Within a pair of tracks for each row it is checked if the assigned clusters are identical. If no GPU counterpart for a CPU track is found, all its clusters differ by definition. Using this criterion the agreement of CPU and GPU is~$99.99976\%$. Fig.~\ref{fig:gp_sizesum} shows that with the improvements the above-mentioned difference in the cluster per track statistic vanishes.

\section{GPU Tracker Performance}

\begin{figure}[ht]
\includegraphics[width=30pc]{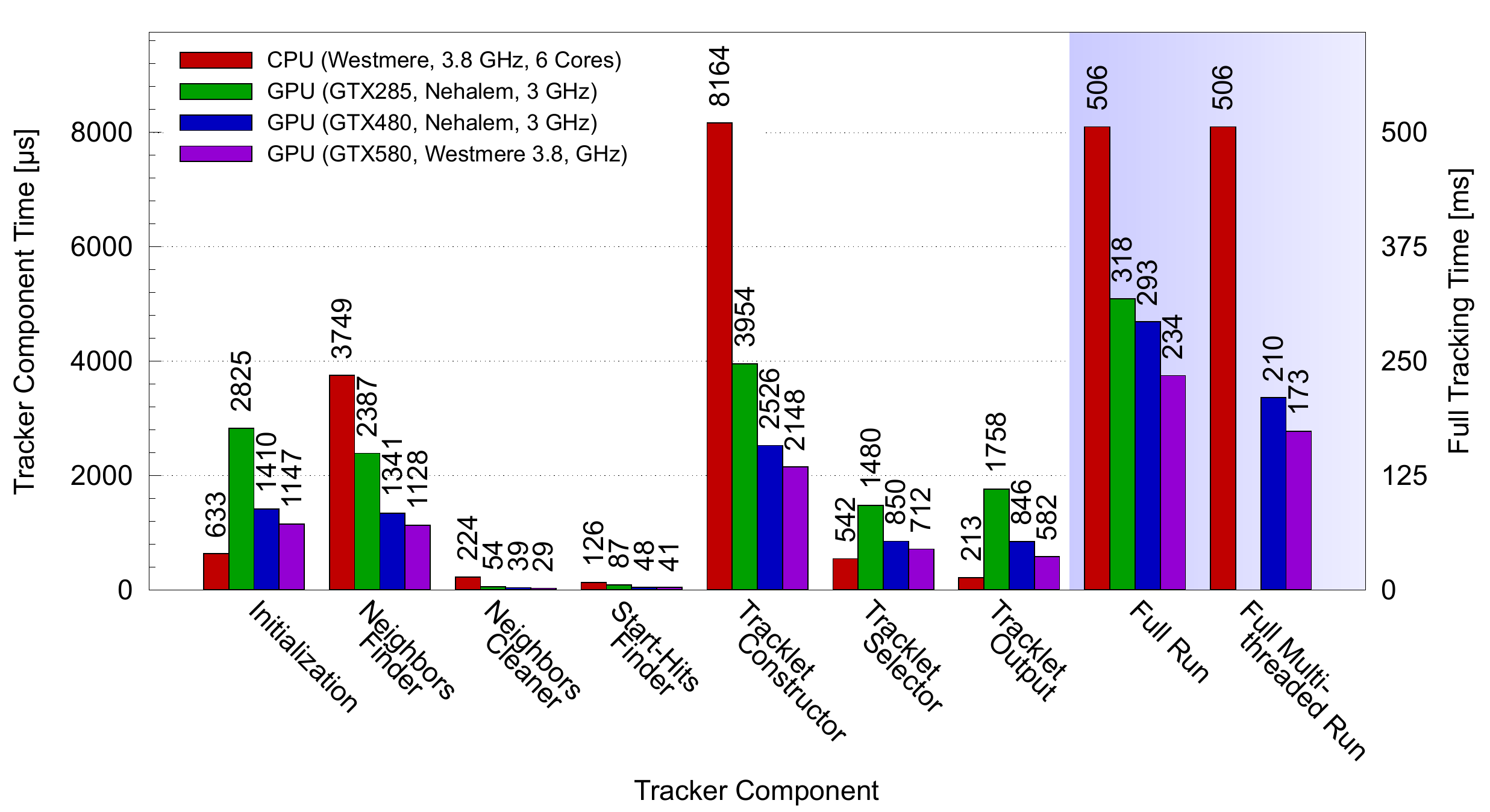}\hspace{2pc}\newline
\begin{minipage}[b]{30pc}\caption{\label{fig:gp_fermi_optimized}GPU Tracker Processing Speed.}
\end{minipage}
\end{figure}

\begin{figure}[ht]
\begin{minipage}{18pc}
\includegraphics[width=18pc]{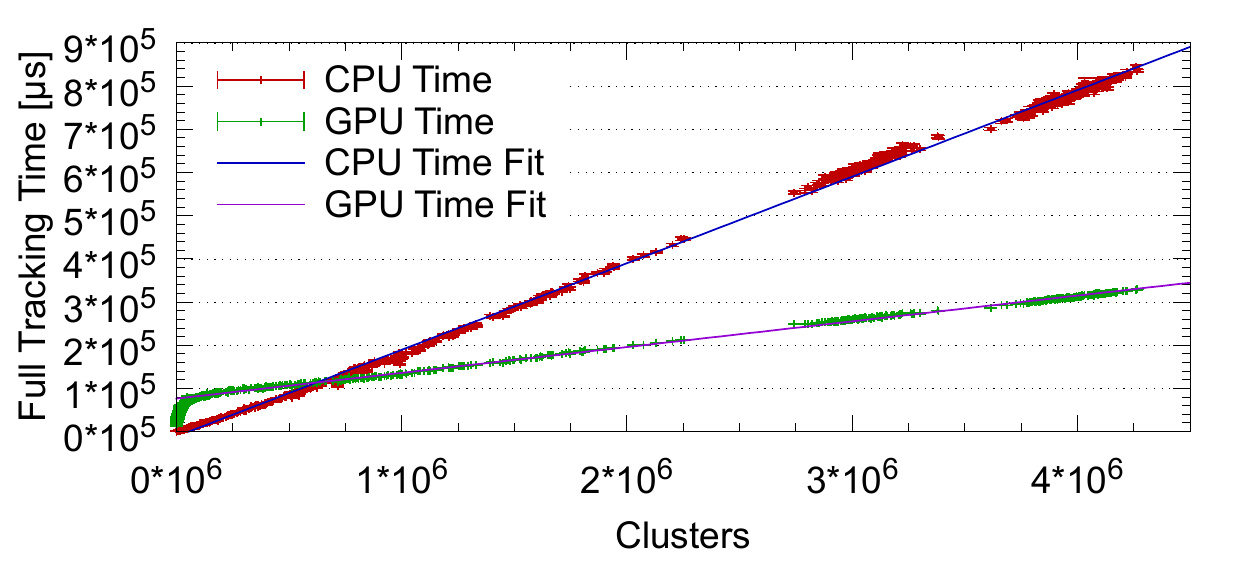}
\caption{\label{fig:gp_eventstat}GPU Tracker Performance Dependency on Input Data Size.}
\end{minipage}\hspace{2pc}%
\begin{minipage}{18pc}
\includegraphics[width=18pc]{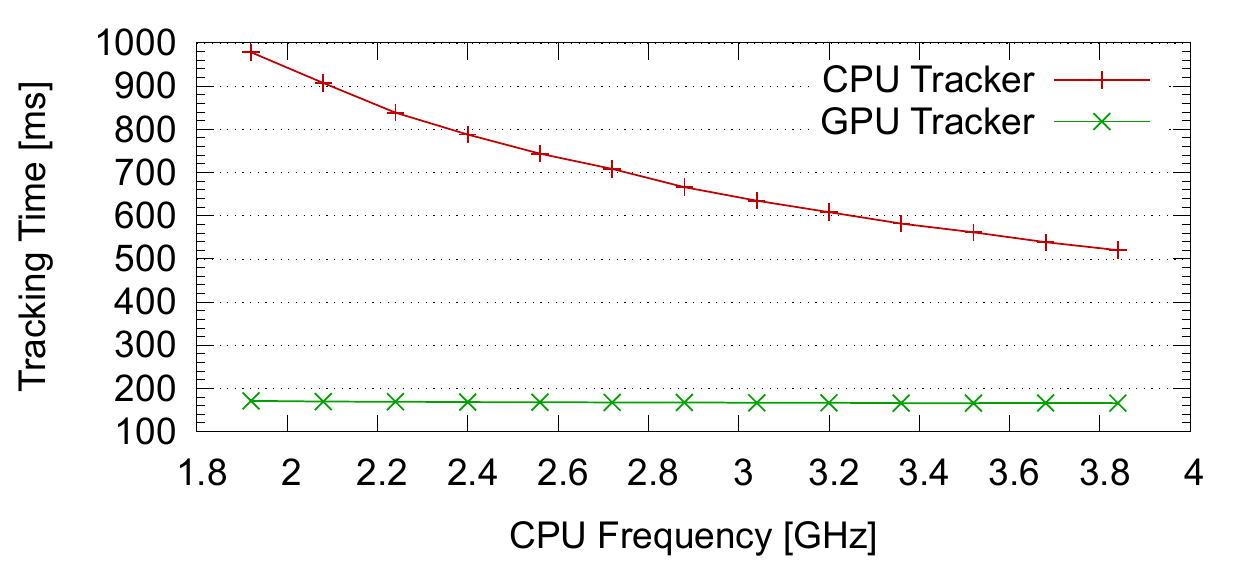}
\caption{\label{fig:gp_cpuscale}GPU Tracker Performance Dependency on CPU Performance.}
\end{minipage}
\end{figure}

Fig.~\ref{fig:gp_fermi_optimized} shows a performance comparison of GPU and CPU tracker for a central heavy ion event. The left part of the diagram presents the processing time of all the tracking steps for one TPC sector and the right part shows the total tracking time for the entire TPC with and without the multi-threaded pipeline. The CPU tracker uses trivial parallelization over the sectors. Since the CPU tracker employs more threads than the multi-threaded GPU tracker, the tasks initialization and tracklet output are faster on CPU than on GPU. Still, using more threads in the GPU tracker is not necessary since the pipelined processing hides the CPU computation time. Fig.~\ref{fig:gp_eventstat} shows that both CPU and GPU tracking time depend linearly on the input data size the GPU tracker having a small offset. Fig.~\ref{fig:gp_cpuscale} shows that with the multithreaded pipeline the actual performance of the CPU is negligible as long as it is fast enough to feed the pipeline.

\section{Global Tracking}

\begin{figure}[ht]
\includegraphics[width=21pc]{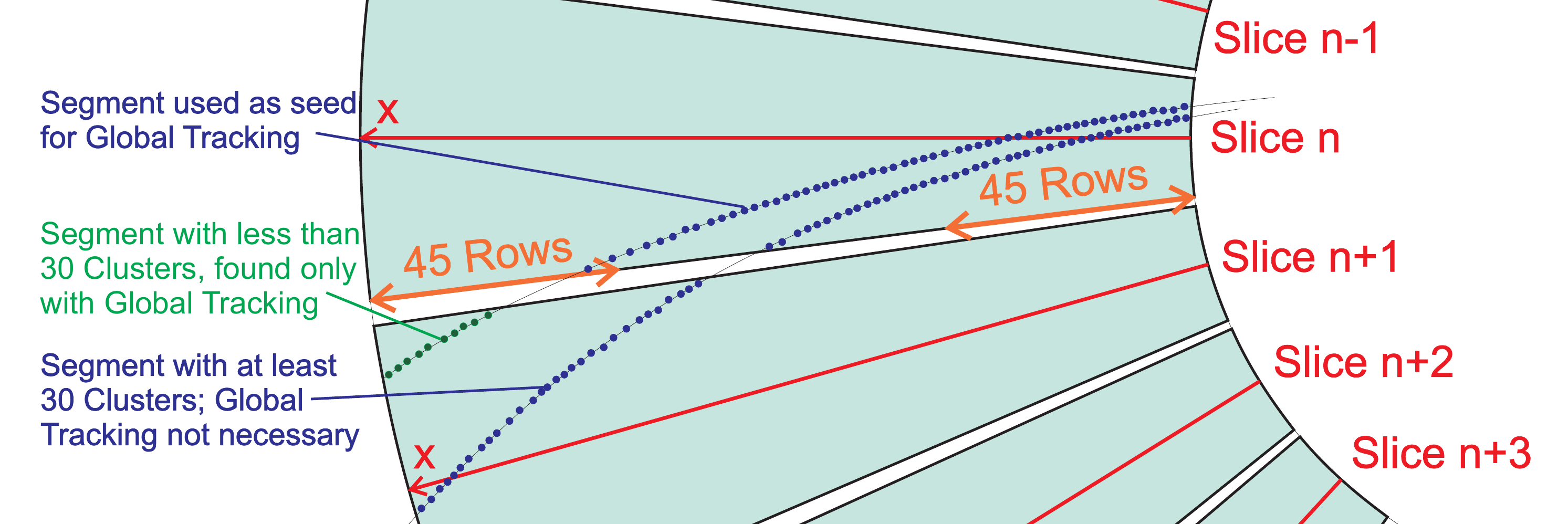}\hspace{2pc}\newline
\begin{minipage}[b]{30pc}\caption{\label{fig:globaltracking}Global versus Local Tracking.}
\end{minipage}
\end{figure}

The sector tracking approach has one inherent shortcoming. The sector tracker only searches for track segments with at least~$30$ clusters. If a track crosses two sectors but the track segment within one of the sectors is very short, the sector tracker does not find it. Hence, an additional step that prolongs track segments is included. Having finished the sector tracking the tracker searches for track segments ending in the innermost or outermost part of the TPC at the edge of a sector. The track parameters are then converted to the coordinate system of the neighboring sectors and the tracklet construction step is continued. Fig.~\ref{fig:globaltracking} visualizes the situation.

Since the prolonged tracking does not find more TPC tracks, tracking efficiency, clone and fake rates do not change with global tracking. In contrast, the resolution improves since the fit now incorporates more clusters.

\section{Tracking during the 2010 and 2011 runs}

\begin{figure}[ht]
\includegraphics[width=21pc]{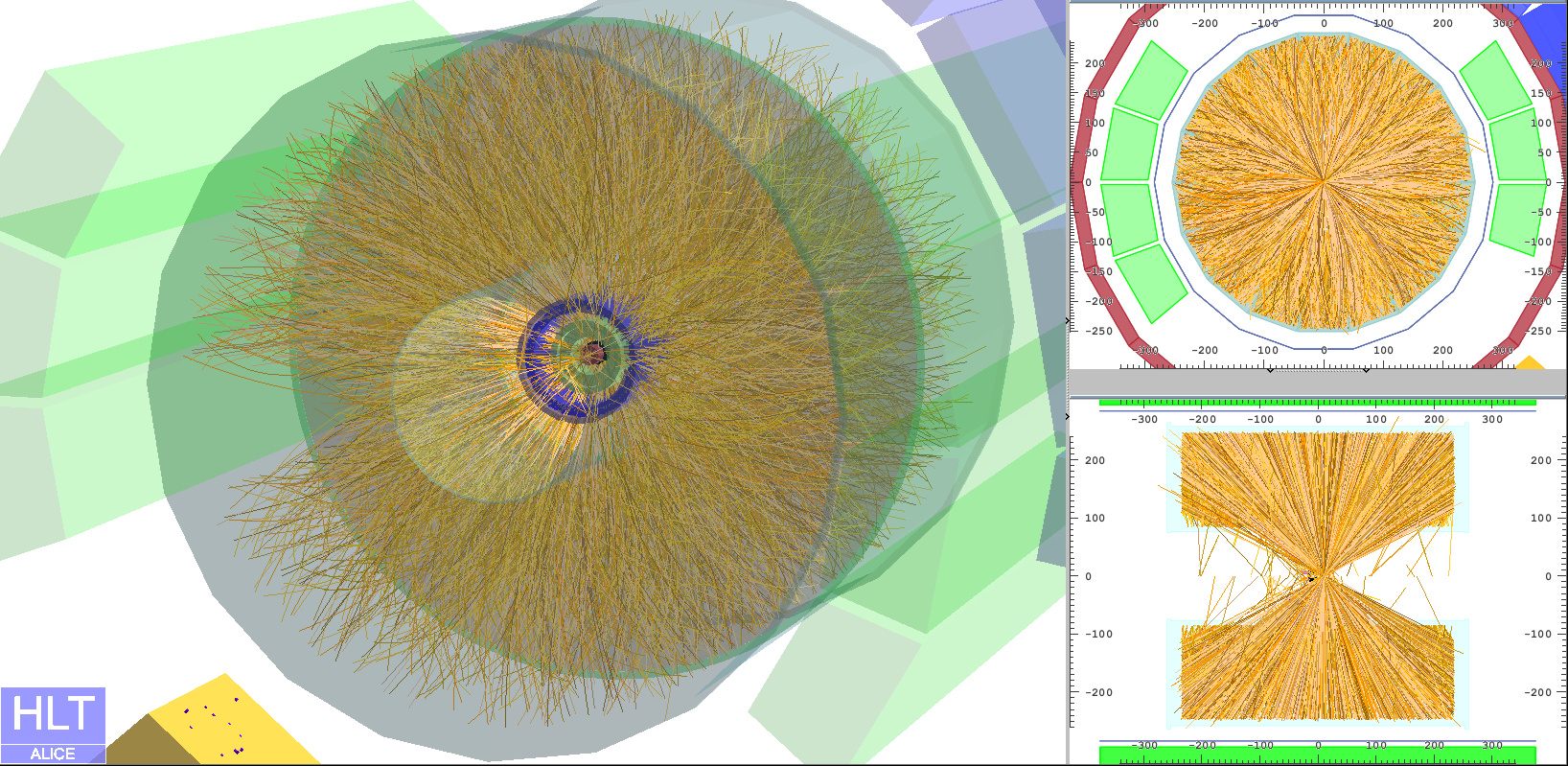}\hspace{2pc}\newline
\begin{minipage}[b]{30pc}\caption{\label{fig:gpuphysics2}Event in first Pb-Pb Physics Run shown in Online Event Display.}
\end{minipage}
\end{figure}

The GPU tracker was operated during the Pb-Pb runs in November 2010 and 2011. Currently, the HLT is equipped with GTX480 GPUs. During this time only one problem emerged. After the TPC had changed the procedure for trip-recovery, it was possible that some sectors were filled with noise, which was not handled properly by the GPU tracker. The problem has been identified and solved. It originated from a synchronization problem between the multiple threads in the pipeline after the GPU runs out of memory due to the immense amount of clusters in a noisy sector. Fig.~\ref{fig:gpuphysics2} shows an image of the online event display during the first run with active GPU tracking.

\section{Conclusion}

The online track reconstruction for the ALICE TPC has been ported to GPUs. The GPU tracker is faster by about a factor of three. It must be noted that the CPU is much more expensive than the GPU so the performance per cost is significantly larger. By and large, the single GPU can perform the tracking equally fast as both CPU chips employing all available cores. The CPU cores, which are still available during GPU tracking, are required for various HLT tasks such as track merging, cluster transformation, vertexing, and so forth. Hence, plugging a GPU in a compute node actually saves the cost of an additional node and of the additional infrastructure required for more nodes. Since the GPUs cost only a fraction of the entire HLT facility, the extra hardware costs for tracking are negligible.

The GPU tracker was proven to be in no way inferior to its CPU counterpart both in terms of efficiency and resolution. It was deployed in November 2010 and proved to be stable in nonstop operation.


\section*{References}
\begin{thebibliography}{9}

\bibitem{alice_technical-proposal}
The ALICE collaboration, ``ALICE - Technical Proposal for A Large Ion Collider Experiment at the CERN LHC,'' CERN, Geneve,
Rep.~CERN-LHCC-95-71;~LHCC-P-3,~1995.

\bibitem{alice_technical-paper}
The ALICE collaboration,
``The ALICE Experiment at the CERN LHC,''
JINST, vol.\ 3, no.\ 08,  Aug.\ 2008.

\bibitem{CARef}
I.~Kisel,
``Event reconstruction in the CBM experiment,''
Nucl.\ Instr.\ and Meth.\ A,\ vol.~566, no.\ 1,\ pp.~85-88, Oct.\ 2006.

\bibitem{bib:tns}
S.~Gorbunov, D.~Rohr et al., ``ALICE HLT High Speed Tracking on GPU'', in 2011 IEEE Transactions on Nuclear Science, vol.~58, no.~4

\bibitem{bib:diploma}
D.~Rohr, ``ALICE TPC Online Tracking on GPGPU based on Kalman Filter``, Diploma Thesis, University of Heidelberg, 2010

\end{thebibliography}
\end{document}